'Oumuamua Is Not a Probe Sent to our Solar System by an Alien Civilization

Ben Zuckerman
Department of Physics & Astronomy, UCLA

ABSTRACT

'Oumuamua, the first known object of extrasolar origin seen to enter our Solar System, has multiple unusual characteristics that, taken together, are very difficult to explain with conventional astronomical entities like asteroids and comets. Consequently, it has been hypothesized that 'Oumuamua is an interstellar probe that was constructed by an alien civilization. We demonstrate that the accomplishments that can be achieved with large space telescopes/interferometers in the alien's planetary system will completely quench any motivation for construction and launch of an 'Oumuamua-like probe. The absence of any such motivation attests that 'Oumuamua is not an alien creation. The existence of large space telescopes has important implications for a range of topics that include interstellar space travel, the Zoo Hypothesis, METI, and UFOs.

## 1. INTRODUCTION

In 2017, the first known object of extrasolar origin -- 'Oumuamua -- transited our Solar System (see Jewitt & Moro-Martin 2020 for a review). 'Oumuamua's properties and trajectory were difficult to explain with what one might call "conventional" models that involve objects such as asteroids and comets that have escaped from main sequence (e.g., Cuk 2018; Jackson et al. 2018; Raymond et al. 2018) or evolved (Hansen & Zuckerman 2017) planetary systems. For reasons given in their paper, Hansen and Zuckerman dubbed such objects "Jurads" to honor their late colleague Michael Jura.

Given that no compelling conventional model to explain 'Oumuamua exists (see Jewitt & Moro-Martin 2020), Bialy & Loeb (2018) suggested that 'Oumuamua is actually an object constructed by a technological, alien, civilization that either (1) passed near the Sun quite by accident, or (2) was intentionally sent by (cosmically nearby) aliens to "probe" our Solar System.

Astronomy is currently in its second golden age (the first followed the invention and use of optical telescopes by Galileo and others of his era). Advances have been driven by the opening of most windows of the electromagnetic spectrum and construction of large new telescopes and sensitive instruments. While even larger ground-based telescopes are planned, these are limited in size and accessible wavelength ranges. The future of astronomy is ultimately in (giant) space telescopes and interferometers, as has been widely recognized for many decades (e.g., Buyakas et al 1979).

Indeed, our nascent technological civilization is already capable of constructing such devices, but has not. At this moment, it is impossible to predict with assurance the consequences of human overpopulation and overconsumption. If and when our species attains a sustainable path, we can be virtually certain that it will include doing astronomy from space. Similarly, one can be virtually certain that any technological civilization that achieves sustainability will construct giant space telescopes.

The relevance of such telescopes to various areas of astronomy has generally been underappreciated. Section 2 describes why their existence renders totally implausible the possibility that 'Oumuamua was deliberately sent by an alien civilization to probe our Solar System. Section 3 touches on the critical concept of motivation and presents a few motivational reasons why 'Oumuamua is not artificial. Section 4 considers some implications of giant space telescopes and of motivations for interstellar space travel, the Zoo Hypothesis, messaging extraterrestrial intelligence (METI), and unidentified flying objects (UFOs).

## 2. 'OUMUAMUA

Bialy & Loeb (2018) propose that "Oumuamua may be a fully operational probe sent intentionally to Earth vicinity by an alien civilization." Their paper contains 22 equations in a few short pages in support of their model of 'Oumuamua as an "extremely thin" interstellar probe. In contrast, other than implicitly assuming that extraterrestrials would want to send such a probe, their paper contains no discussion of the alien's motivations. More than 50 years ago, physicist Freeman Dyson understood that ignoring the question of motivation would be a huge mistake:

"The problem of interstellar travel is a problem of motivation and not of physics"
(in "Essays in Honor of Hans A. Bethe", July 1966).

It is possible to envision plausible motivations for sending a probe from one planetary system to another; examples are presented in Zuckerman (1981 & 2019) and Hansen & Zuckerman (2021). However, as argued in this section and in section 3, no plausible motivation exists for a probe that has the characteristics of 'Oumuamua as described by Bialy and Loeb (2018).

In an article entitled "'Oumuamua is not Artificial", Katz (2021) presents various reasons to justify his title and notes that "a flyby is an inefficient way to collect data". At the measured velocity at infinity of 'Oumuamua relative to the Sun, 26 km/s, it would take of order 50,000 years to travel between the nearest star (1.3 pc away) and the Sun. Going out to a distance of 10 pc would encompass only 357 main sequence stars (Henry et al. 2018); from 10 pc it would take about 400,000 years to arrive here at a velocity of 26 km/s.

Thus, any technological civilization that could have sent the 'Oumuamua "probe" must have anticipated that it would still be in existence (at least) 50,000 years later, and very likely far longer. Our species has had technology for only a few hundred years and the capability to go into space for only about 50 years. In this brief time interval we have developed the technological potential to build space telescopes with a variety of designs to detect and characterize nearby extrasolar rocky planets that are potentially capable of supporting life.

Zuckerman (2019) discusses the relevance of such telescopes to demonstrate how unlikely it is that any technological life exists in orbit around any nearby star. But our aim here is quite different. Should a nearby technological civilization actually exist, then, as we show below, its giant space telescopes would completely vitiate any motivation for construction of the type of directed probe that 'Oumuamua represents in the Bialy & Loeb (2018) picture.

Earth-based telescopes have played an absolutely essential role in our exploration of our Solar System by spacecraft. NASA's Perseverance is only the most recent of a string of spacecraft sent to explore Mars, rather than, say, Mercury or Venus; this is because our telescopes have long provided ample *motivation* to go to Mars above all other planets. Similar considerations, but for space-based telescopes, hold for exploration of extrasolar planets (e.g., Zuckerman 2019). Forty years ago, Zuckerman (1981) pointed out the deep connections between space telescopes and interstellar probes in a paper titled "Space Telescopes, Interstellar Probes, and Directed Panspermia".

That article noted that "Large space telescopes should permit detailed study of nearby planetary systems. It will be possible to obtain orbital and spectroscopic data of a quality comparable to that now obtainable from ground-based telescopic study of planets in our Solar System". Below, we explicitly compare the capabilities of space telescopes in a system located parsecs away with those of 'Oumuamua as it passed through our planetary system.

First generation planet-hunting space telescopes such as NASA's Terrestrial Planet Finder ("TPF") or ESA's DARWIN (see Appendix A in Zuckerman 2019 for a description of TPF and DARWIN), would be able to discover "living worlds" such as Earth. This would be accomplished through identification of "biomolecules" in the planet's atmosphere. Once such a world is found, then surely larger, more powerful, space telescopes would be constructed to resolve the planet's surface well enough to reveal continents and oceans, the spectrum of "vegetation", and variations in these as a function of time, both in the short-term (annual seasons) and the long-term (ice ages, or whatever).

In the mere 50 years of the space age, construction of TPF or DARWIN could easily have already begun, but has not because politicians have decided to spend the money on something else – for example, the hundreds of billions of dollars spent annually on the military. Presumably any technological species that sends out probes that will take at least 50,000 years to reach their destinations has figured out a way to live sustainably in its biosphere.

The nearest stars have been the Sun's neighbors for millions of years. Thus, the technological creatures on our hypothetical nearby world would have had more than ample time to use their equivalent of an advanced version of TPF to discover and study

Earth as a living world well before they would consider launching an 'Oumuamua. This ordering of events would be completely analogous to our telescopic investigation of the Solar System before we sent any rockets into interplanetary space. Given what we now know about our own and extrasolar planetary systems, it is fair to say that the only remarkable aspect of our planetary system is Earth itself. Thus, a first question one might ask is: if the probe 'Oumuamua was sent to study our planetary system, then why was its closest approach to the most interesting planet not much closer than 0.2 au?

Let's look at 'Oumuamua in a little detail and consider its potential capabilities. With dimensions of about 10 m by 100 m, the largest telescope that 'Oumuamua could reasonably possess would have a diameter of ~10 m. Could the fragile spacecraft that Bialy & Loeb (2018) describe even carry such an instrument and associated equipment? And would everything still work after an interstellar voyage of at least 50,000 years? Nonetheless, continuing, we can calculate the spatial resolution achievable with a 10 m telescope a distance of one au from its observational target. One au is motivated by the fact that 'Oumuamua traveled through the Solar System at a speed of about 5.5 au/year and thus spent a few months at about this distance from Earth around the time of its closest approach to Earth.

For comparison, one can calculate the spatial resolution for objects in our planetary system that could be achieved by a space-based interferometric array in a planetary system located, for example, 5 pc (= $10^6$ au) from the Sun. For a modest array dimension similar to the Earth-Moon separation, the spatial resolution of the array would better, by more than an order of magnitude, that of the hypothetical 10-m telescope carried by 'Oumuamua when it was one au from Earth. With 10,000 or more years at their disposal, no doubt the aliens could and would construct even much larger arrays for all kinds of studies of the Universe.

In any event, an array could study Earth for tens of 1000s of years, in comparison with the few month flyby of 'Oumuamua as it zips through the Solar System. As Jonathan Katz noted above, "a flyby is an inefficient way to collect data". Given the remarkable capabilities of a large space interferometer located in the alien's planetary system, the only plausible probe the aliens would want to send to our Solar System would be one that would remain here for a long time, probably in close orbit around Earth. Such a complex probe is surely not anything that looks like 'Oumuamua.

Given that a directed alien probe is a completely implausible explanation for 'Oumuamua, one must fall back on explanation option (1) in the Introduction, an accidental encounter of an artificial probe with our solar system. This explanation also totally fails Freeman Dyson's motivation test for interstellar travel. Jewitt et al (2017) estimate the frequency of arrival of 'Oumuamua-like objects into the Solar System as about three per day! Jewitt & Moro-Martin (2020) mention the implications of this arrival rate:

"What, then, does this frequency suggest about the origin of 'Oumuamua? Aliens might be capable of sending a Saturn V–size rocket or a large piece of Mylar-like material across the galaxy and through our solar system, but why would they send so many? Even more astonishing, if we extrapolate our analysis from the solar system to the whole of the Milky Way, we find that there must be $10^{24}$ to $10^{25}$ similar objects in our galaxy. It is hard to believe that an extraterrestrial civilization would have the capacity to flood the galaxy with so much space junk, and it is even more difficult to see why it would do so."

3. ON MOTIVATION

Section 2 presents a few reasons why 'Oumuamua is not a probe sent to our solar system by an alien civilization. In the present section we mention additional motivational arguments as to why 'Oumuamua is not artificial.

For travel between star systems a variety of propulsion systems – that yield a wide range of velocities at infinity following escape from the gravitational field of a star -- have been suggested (e.g., Dyson 1995, Singer 1995, Crawford 2018). The time required for the capabilities of a technological society to advance from a given method of propulsion to a faster one is surely much less than the interstellar travel time of 'Oumuamua to reach our solar system (as noted above, an absolute minimum of ~50,000 years, and no doubt substantially longer). The putative beings that sent 'Oumuamua must have been aware of these relative time intervals – this would have quenched any desire to send 'Oumuamua, at basically a snail's pace, toward the Sun.

An example of a potential (rapid) advance in propulsion systems would be the step from chemical to nuclear fission rockets. We already know how to generate and control fission energy in stationary and in moving (submarines) applications. A comparison of chemical and fission propulsion appears in some papers on Project Icarus (a theoretical design study for an interstellar probe). Freeland (2013) gives theoretical maximums for chemical and uranium fuel: "The highly-optimized LOX/LH2 chemical fuel commonly used on today's spacecraft provides energy output of approximately 13 MJ/kg. Fission of uranium provides energy output of around 60 million MJ/kg (60 TJ/kg)."

What motivation would any being have for launching a mission whose outcome would not be known for of order 100,000 years or more? One might point to the "Arecibo radio message" that Frank Drake and Carl Sagan launched in 1974 toward the globular star cluster M13, ~25,000 light years from Earth.
(see https:en.wikipedia.org/wiki/Arecibo_mesage)
But the difference in time, energy, and expense between this radio signal and 'Oumuamua is so gigantic as to render such a comparison meaningless.

One might argue that we cannot know what would motivate a being far in advance of ourselves. But 'Oumuamua is traveling at only 26 km/s with respect to the Sun. Our current interstellar travel capability is represented by the Voyagers; these relied on gravity assists off the giant planets. The Voyagers have achieved effective terminal velocities of ~20 km/s. So when 'Oumuamua was launched, the responsible beings were, arguably, technologically not much different from us, and their motivations for interstellar travel might well not have differed from ours.

From the time of the caveman to the present, advances in technology have been truly incredible. By contrast, the things that motivate our species have changed but little. The caveman's principal motivations were short term – what to do when the saber-toothed tiger appeared at the mouth of one's cave?  In this respect, we clearly have not changed over the millennia. This is why, as economists might say, we continue to discount the future even as we march toward catastrophes; rapid climate change and the COVID-19 pandemic probably represent just the tip of an approaching iceberg.  Given the slow evolution of motivation, there is no reason to expect that the motivations of an alien species will differ so much from our own that, somehow, the launching of 'Oumuamua would be logically plausible to them while logically obscure to us.

In sections 2 and 3 we present motivational arguments as to why 'Oumuamua is not a probe sent to our solar system by an alien civilization.   As noted two paragraphs above, 'Oumuamua travels at a pedestrian (plodding) speed through the Galaxy.  Arguably, our species will have the capability to build much speedier nuclear fission rockets in the, cosmically speaking, very near future.  Thus, one might reasonably argue that the species that launched 'Oumuamua was only marginally more technologically advanced than our own and thus would be unlikely to have evolved motivational qualities that differ much, if any, from our own.

Alternatively, ignoring 'Oumuamua's dawdling speed, we can assume that the species that launched it is technologically far in advance of our own.  A fundamental issue then is what might motivate a technological civilization that might, for example, be governed by artificial intelligence (AI) rather than a carbon-based intelligence that has evolved through eons of biological time.   Surely a technological species that has persisted for times of order 100,000 years or more has learned how to deal with the multitude of problems that confront our species at the dawn of the age of technology.

If so, then one might reasonably ask: "what might motivate a species that might be guided by AI and has achieved sustainability"?   Because no one can know the answer to such questions with certainty, the arguments presented here are just that – arguments – and not an absolute answer to the question of what is 'Oumuamua.  That said, these arguments have the virtue of satisfying Occam's razor.

## 4.  THE ZOO HYPOTHESIS, METI, AND UFOs

Motivation is also a major consideration for some paradigms related to extraterrestrial technology.   An idea that first appeared in the scientific literature in 1973 is the Zoo Hypothesis (Ball 1973), although it was predated in the science fiction literature (see, e.g., Webb 2002, chapters 5 & 6).  The Zoo idea is that "They" are indeed out there but are purposely remaining hidden from our view while they keep us under surveillance.  In essence we live in a zoo or wilderness preserve.  Thus the Zoo Hypothesis assumes that we humans occupy center stage, otherwise what would They be hiding from?   This solipsistic point of view violates the Copernican principle.  Ball's paper continues to attract attention, being cited more during the past decade than during the 35 years following its publication.

Our species is curious about life in the Universe and we might reasonably expect other technological beings to have a similar interest. As noted in Zuckerman (2019), any nearby technological civilization would have used their space telescopes to discover that Earth is a noteworthy place long before we humans ever appeared on the scene. These beings would have had eons of time to send complex probes (or perhaps themselves) to our solar system to study life on Earth in ways not possible from the distance of their own planetary system.

Once in our planetary system, they would have no reason not to set up shop in the most obvious ways, because what is the point of hiding from microbes, dinosaurs, fish, and trees? By the time technological humans finally appear during the past few 100 years, the alien's robotic devices or the aliens themselves -- even if not on Earth itself -- would be sprawling about and very much in plain sight.

Similar considerations apply to the shy, furtive, UFOs that never seem to go away. In the decades that followed WWII, fuzzy "pictures" of unidentified objects often appeared in the media. But now that half the world's population – billions of persons – possess cell phones and are amateur photographers, we still do not have a single convincing picture of an alien craft of any sort. When a real object, such as a bright meteor, appears multiple cell phones often record its existence. UFO sightings have been reported from many countries including Canada, France, Chile, Japan, China and other countries in Europe and South America. Chile and France have sponsored official investigations into UFOs.

In 2021 the U.S. Office of the Director of National Intelligence issued a "Preliminary Assessment: Unidentified Aerial Phenomena". To quote the Los Angeles Times: this "Long-awaited UFO report has few answers." This 2021 assessment focuses on reports, sometimes by military pilots, of fleeting apparitions, but with no more evidence of an extraterrestrial origin than has been supplied by the billions of cell phone cameras. In any event, one must again ask: what is the motivation for the aliens to play hide and seek with us?

Following WWII there was sufficient interest in unidentified flying objects that the U. S. Air Force conducted studies for about two decades; these were eventually gathered together in Project Blue Book. The Air Force contracted with the University of Colorado to evaluate material contained in the Blue Book and, also, obtained from civilian UFO groups. A resulting report, the "Scientific Study of Unidentified Flying Objects" was headed by physicist Edward Condon and has come to be known informally as the Condon Report.

The report concluded that continued study of the UFO phenomenon was not scientifically warranted and it supplied no support for the idea that UFOs have anything to do with extraterrestrial visitors. This report was submitted to the National Academy of Sciences, which concurred with its conclusions: (http://project1947.com/shg/articles/nascu.html)

Nonetheless, the media continues to keep the extraterrestrials idea alive; one consequence is the 2021 report referred to above.

Perhaps the fact that we are an aggressive, destructive species would motivate extraterrestrials to stay hidden. A similar consideration sometimes appears in METI -- messaging extraterrestrial intelligence – but in this picture it is the aliens who are the aggressive, nasty, creatures. Many persons, including some very famous ones, believe that we should do all we can to remain invisible to extraterrestrials for fear they will come to Earth to conquer or wipe us out (e.g., Haqq-Misra et al. 2013).

Not only is this an extreme solipsistic (and simplistic) concern, but it projects current human behavior onto beings in a civilization that has existed for far longer than ours and who have, no doubt, learned how to live sustainably with their available resources. As noted above, any technological civilization that might be close enough to care about earth life, would have been monitoring Earth for eons with its space telescopes and would have had far more appealing reasons to come here other than to *wipe us out*.

## 5. CONCLUSIONS

The existence and capabilities of large space telescopes cannot be ignored when one endeavors to understand motivations for interstellar travel. In contrast to the multiple motivations that technological beings would have for construction of such telescopes, for example, as given in Zuckerman (1981 & 2019), we demonstrate that there exists no plausible reason why a technological civilization would build and launch 'Oumuamua type probes of the sort described by Bialy & Loeb (2018).

The fleeting capabilities of any such flyby probe are vastly inferior to the power of space telescopes operational for eons of time in the interplanetary space of the alien civilization. Although it has proven difficult for astronomers to construct a compelling picture of 'Oumuamua based on conventional astronomical objects, the discussion in the present paper demonstrates the logical implausibility of an explanation that involves flybys of alien probes.

What then is 'Oumuamua? The answer remains elusive (see, e.g., Jewitt & Seligman, 2023). Many papers that present models for 'Oumuamua have appeared. However, none stand out as much more likely to be correct when compared to the others.

We also consider the importance of motivations and/or the potential role of space telescopes when evaluating the plausibility of visitations to our solar system by shy, secretive, or hostile alien beings.

**Note Added in Proof**

The 2021 ''Preliminary Assessment: Unidentified Aerial Phenomena'' mentioned in Section 4 has now been followed up by NASA which has commissioned a 16 person panel to investigate ''unidentified aerial phenomena'' (UAP). The panel's report will be due in 2023. It will look at previously collected UAP observations (only unclassified ones, not classified data), focusing on how they could be better organized and analyzed in the future to shed more light on mysterious sky sights. Such sights seen over, for example, the Pacific Ocean are more likely to be something akin to high velocity Chinese drones than visitors from other worlds. The NASA announcement can be found at this website: https://science.nasa.gov/uap

**Acknowledgements**

We thank Robert Sheaffer, Brad Hansen, and David Jewitt for helpful comments.